\documentclass[twocolumn,english,aps,prl,groupedaddress,superscriptaddress]{revtex4}

\usepackage[pdftex]{graphicx} 
\usepackage{xcolor} 
\usepackage{soul} 
\usepackage{amsmath,amsfonts,mathrsfs,amsbsy,bm,babel}
\usepackage{dcolumn}
\usepackage{bm}
\usepackage[utf8x]{inputenc}
\usepackage{textcomp}
\begin{document}
\title{Magnetic excitations and anomalous spin wave broadening in multiferroic FeV$_{2}$O$_{4}$}
 
\author{Qiang Zhang}
\affiliation{Ames Laboratory, Ames, IA, 50011, USA}
\affiliation{Department of Physics and Astronomy, Iowa State University, Ames, IA, 50011, USA}

\author{Mehmet Ramazanoglu}
\affiliation{Ames Laboratory, Ames, IA, 50011, USA}
\affiliation{Department of Physics and Astronomy, Iowa State University, Ames, IA, 50011, USA}

\author{Songxue Chi}
\affiliation{Oak Ridge National Laboratory,Oak Ridge,Tennessee 37831, USA }

\author{Yong Liu}
\affiliation{Ames Laboratory, Ames, IA, 50011, USA}

\author{Thomas. A. Lograsso}
\affiliation{Ames Laboratory, Ames, IA, 50011, USA}
\affiliation{Division of Materials Sciences and Engineering, Iowa State University, Ames, Iowa 50011, USA}

\author{David Vaknin}
\affiliation{Ames Laboratory, Ames, IA, 50011, USA}
\affiliation{Department of Physics and Astronomy, Iowa State University, Ames, IA, 50011, USA}

\date{\today}

\begin{abstract}
 We report on the different roles of two orbital-active Fe$^{2+}$ at the A 
site and V$^{3+}$ at the B site in the 
magnetic excitations and on the anomalous spin wave broadening in 
FeV$_{2}$O$_{4}$. FeV$_{2}$O$_{4}$ exhibits three structural transitions and 
successive paramagnetic (PM)-collinear ferrimagnetic (CFI)-noncollinear 
ferrimagnetic (NCFI)  transitions. The high-temperature tetragonal/PM 
-orthorhombic/CFI transition is accompanied by the appearance of an 
energy gap with a high magnitude in the magnetic excitations due to strong spin-orbit coupling induced anisotropy at the Fe$^{2+}$ 
site. While there is no measurable increase in the energy gap from the 
orbital ordering of V$^{3+}$ at the orthorhombic/CFI-
 tetragonal/NCFI transition, anomalous spin wave broadening is observed in the orthorhombic/CFI state due to V$^{3+}$ spin fluctuations at the B site. 
The spin wave broadening is also observed at the zone boundary without 
softening, 
which is discussed in terms of magnon-phonon coupling. 

\end{abstract}

 \maketitle
     
     Understanding the orbital degrees of freedom and their coupling with spin 
and lattice degrees of freedom has emerged as a forefront topic in modern 
condensed-matter physics as these coupled degrees of freedom play a central 
role in inducing novel phenomena\cite{Tokura2000}. Vanadium spinel 
oxides with formula
 AV$_{2}$O$_{4}$ are ideal systems to study the 
 orbital ordering (OO) by virtue of the fact that the 3d$^{2}$ high-spin configuration of V$^{3+}$  is accommodated in the triply degenerate t$_{2g}$ states rendering it with orbital
degrees of freedom. For a non-magnetic occupancy of the A site by a divalent ion such as Zn, Mg and Cd \cite{Niziol1973}, there is usually a structural transition from cubic to tetragonal, followed by a magnetic ordering 
at a lower temperature. Replacing A by a magnetic ion Mn$^{2+}$ in a 3d$^{5}$ high spin configuration without orbital degrees of freedom leads to different magnetic transitions although
there exists a similar structural transition: 
a paramagnetic (PM)-collinear ferrimagnetic (CFI) in the same cubic symmetry at $T_{N1}\approx 56 $ K, followed by a CFI-noncollinear ferrimagnetic (NCFI) transition accompanied by the 
cubic-tetragonal structural transition at \textit{T}$_{N2}\approx$ 53 K. In 
FeV$_{2}$O$_{4}$, 
the A-site Fe$^{2+}$ with a high-spin 3d$^{6}$ configuration and three electrons
in the doubly degenerate \textit{e }states gives rise to 
orbital degrees of freedom of Fe$^{2+}$\cite{Katsufuji2008, Sarkar2011}. 
FeV$_{2}$O$_{4}$ exhibits similar PM-CFI-NCFI transitions as in 
MnV$_{2}$O$_{4}$, but the competition or cooperation between two 
orbital-active Fe$^{2+}$ and V$^{3+}$ leads to controversial
 three\cite{Zhang2012,MacDougall2012} or four structural 
transitions\cite{Katsufuji2008}. 
Previous  investigations have focused on
 the orbital ordering of V$^{3+}$ at the B site and its effect on the cubic-tetragonal transition and the magnetic excitations in 
$A$V$_{2}$O$_{4}$ ($A=$ Zn, Mg, Cd and 
Mn).\cite{Perkins2007,Matteo2005,Tsunetsugu2003,Garlea2008,Nanguneri2012}
FeV$_{2}$O$_{4}$ provides a good candidate to investigate the roles of 
orbital orderings on both sites. Recent discovery of multiferroicity in FeV$_{2}$O$_{4}$ 
\cite{Zhang2012,Liu2012}, in coexistence of ferroelectricity and 
noncollinear ferrimagnetism in contrast to  the antiferromagnetism in most of 
the multiferroics, further motivates us to focus on this system. The 
ferroelectricity is not found in the collinear ferrimagnetic phase and only 
emerges in the noncollinear ferrimagnetic phase. It is of 
 interest to compare the spin dynamics 
\cite{Lynn2000} in these two distinct magnetic phases, and also to figure out 
the source of the spin frustration in FeV$_{2}$O$_{4}$ since spin frustration is 
usually related to the 
appearance of the ferroelectricity in various
noncollinear magnetic phases \cite{Arima2011,Kim2011}. Here, we report
 elastic and inelastic neutron scattering results on high-quality FeV$_{2}$O$_{4}$ single crystal.

\begin{figure} \centering \includegraphics [width = 1.\linewidth] {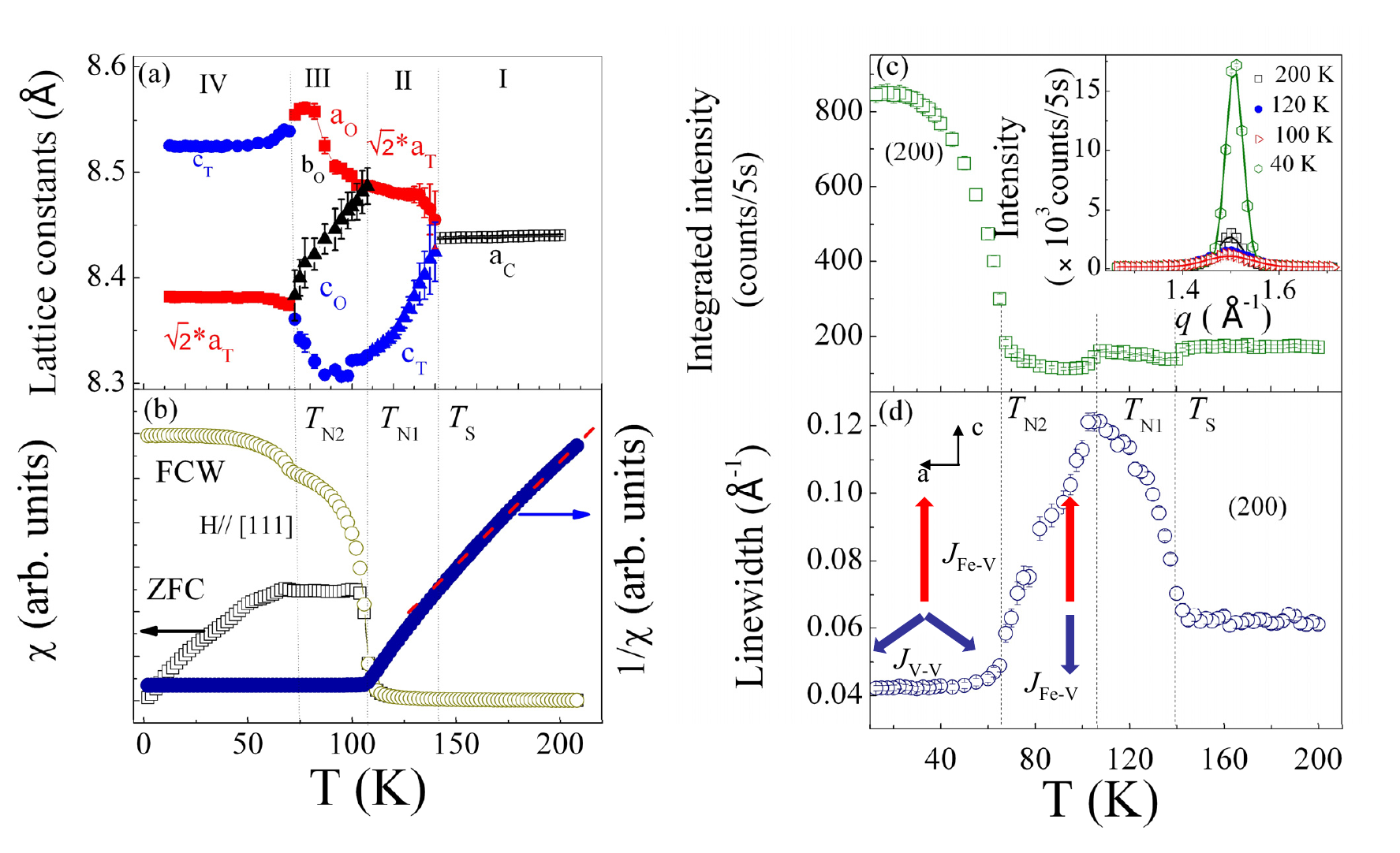}
\caption{(color online) (a) Temperature dependence of the lattice parameters in FeV$_{2}$O$_{4}$. (b) Temperature dependence of the DC susceptibility (squares) after zero-field-cooled 
and field-cooled  with H//[111] axis in FeV$_{2}$O$_{4}$. The solid circles show the inverse susceptibility. 
Temperature dependence of (c) the integrated intensity and (d) linewidth of (200) peak. The inset of (c) shows representative \textit{q} scans of the (200) peak.
The solid lines are fits to a Gaussian function. The schematic magnetic structures projected on the \textit{ac} plane with the main magnetic interactions below/above $T_{N2}$  are also shown in (d). The  long arrows and short  arrows represent Fe$^{2+}$ and V$^{3+}$ spins, respectively. Thre dashed lines mark  the three transitions.}
\label{fig:susceptibility} 
\end{figure}

 The experimental details and the schematic evolution of the structure based on the splittings of the (400) Bragg peak in the cubic setting are provided in the \textgravedbl Supplemental Material\textacutedbl. The obtained lattice constants as a function of temperature shown
 in Fig.\ \ref{fig:susceptibility} (a) indicate that there are three structural transitions:
cubic-high-temperature (HT) tetragonal ($c<a$) at $T_{S}=140$ K, HT 
tetragonal-orthorhombic at $T_{N1}=110$ K, orthorhombic-low-temperature (LT)
 tetragonal at $T_{N2}=70$ K and no other tetragonal-orthorhombic structural 
transition at $\approx$ 35 K as reported by 
Katsufuji \textit{et al.} \cite{Katsufuji2008} but not by others\cite{MacDougall2012,Nii2012}.     
   Figure\ \ref{fig:susceptibility}(b) shows the temperature dependence of the DC susceptibility after zero-field-cooling and field-cooling (FC) in magnetic field of 
1000 Oe parallel to the [111]. Below $T_{N1}$, a rapid increase in the susceptibility is ascribed to 
the PM to CFI ordering where the Fe$^{2+}$ moments are parallel to [001] and the V$^{3+}$ moments are antiparallel to Fe$^{2+}$ moments via antiferromagnetic coupling \textit{J}$_{\rm Fe-V}$ \cite{Zhang2012,MacDougall2012}.
 Another jump in the FC susceptibility below $T_{N2}$ results from CFI to NCFI 
transition due to the V$^{3+}$ canting \cite{Zhang2012,MacDougall2012} along any 
of $<$111$>$ directions.\cite{MacDougall2012}.
 The schematic CFI and NCFI magnetic 
structures projected on the \textit{ac} plane with the main magnetic 
interactions\cite{MacDougall2012, Zhang2012} are shown in Fig.\ \ref{fig:susceptibility}(d).
 The inverse susceptibility shows 
a deviation below $T_{S}$, indicating a magnetoelastic coupling \cite{Katsufuji2008} at $T_{S}$. We note that FeCr$_{2}$O$_{4}$ with only orbital-active Fe$^{2+}$
exhibits similar  PM cubic-PM tetragonal($c<a$)-CFI orthorhombic transitions\cite{Bordacs2009}  without the lowest one
 at $T_{N2}$. Thus, the two transitions at $T_{S}$ and $T_{N1}$ in FeV$_{2}$O$_{4}$
are mainly ascribed to the involvement of orbital-active Fe$^{2+}$.\cite{Nii2012} The LT tetragonal phase with $c >a $ in FeV$_{2}$O$_{4}$ is unique in all the vanadium spinel oxides, 
suggesting both orbital-active Fe$^{2+}$  and V$^{3+}$  are necessary\cite{Katsufuji2008} to induce the third
structural transition at $T_{N2}$.

      Figure\ \ref{fig:susceptibility} (c) and (d) show the temperature dependence of integrated intensity and peak line-width of the high-symmetry-cubic-forbidden (200) reflection 
in FeV$_{2}$O$_{4}$.  The peak is present at all measured temperatures and exhibits anomalies at the three transitions. 
We note that the observed (200) is not due to $\lambda$/2 leakage as the PG filters remove this higher order wavelength  to better than one part
 in $3\times10^{6}$ as measured on the nuclear (220) Bragg peak and the forbidden (110) at 200 K.   
The observed (200) peak above $T_S$ at 200 K has a pure structural origin due to anisotropy of the local environment around the transition-metal atoms with
 no contribution from charge-ordering (CO) or OO, as discussed in other spinels, such as $A$Fe$_{2}$O$_{4}$ ($A=$ Mn, Co and Fe)\cite{Subias2004}. Whereas weak anomalies at $T_{S}$ and $T_{N1}$ in the intensities and linewidths are present,
 the (200) reflection with higher  intensity and  sharper peak below $T_{N2}$ is mainly magnetic in origin, which confirms the 
occurrence of V$^{3+}$ spin canting as depicted in Fig.\ \ref{fig:susceptibility} (d).

\begin{figure} \centering \includegraphics [width = 1.\linewidth] {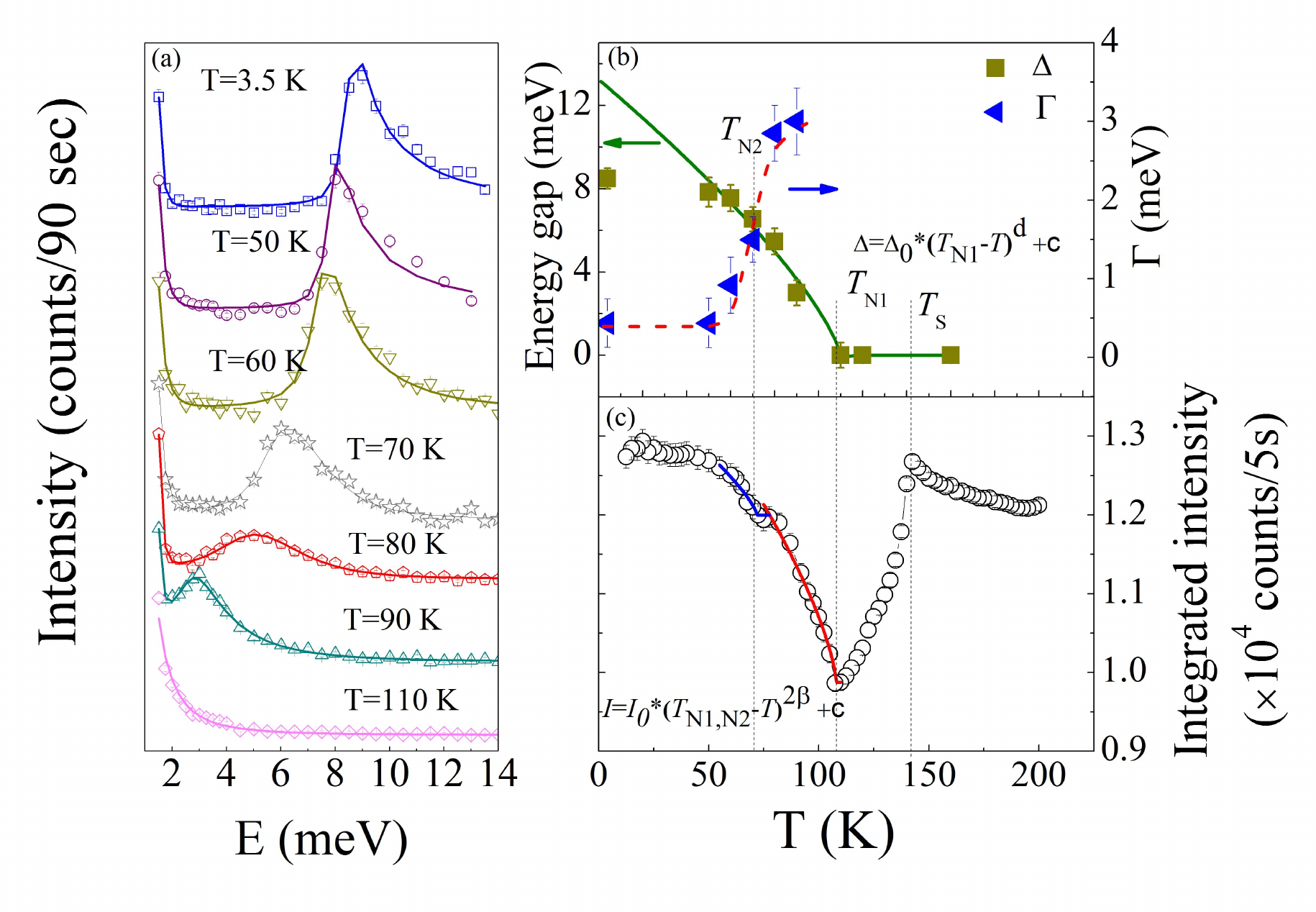}
\caption{(color online) (a) Constant-\textbf{Q} energy scans measured at the zone center (220) at various temperatures. The solid lines are fits using the model described in the text.(b)
Temperature dependence of (b) the energy gap, damping factor $\Gamma$, and (c) the integrated intensity of (220) Bragg peak. The solid lines are fits to
 the data (see text for more details).}
\label{fig:Escan} 
\end{figure}

      Constant-\textbf{Q} energy scans were measured at the zone center (220) at various temperatures and at various \textbf{Q}'s along [H H 0] 
at CFI (90K) and NCFI (3.5 K) phases. As shown in Fig.\ \ref{fig:Escan} (a), 
a clear energy gap $\approx 8$ meV at (220) in low-$E$ region is observed at 3.5 K and 
the gap drops smoothly with increasing temperature. In the damped simple 
harmonic oscillator approximation \cite{Ramazanoglu2013, Pratt2010, Ye2007}, 
the neutron scattering cross-section is given by :
%
%
\begin{equation}
  \frac{d^{2}\sigma}{d\Omega dE}(\textbf{q},E) \propto  \frac{A_{q} \Gamma E}   {[E^{2}-(\hbar \omega(\textbf{q}))^{2}]^{2}+\Gamma^2E^2}    (1-e^{-E/kT})^{-1}
  \label{eq:DampSW}
\end{equation}

  where A$_{q}$ is \textit{q}-dependent intensity, $\Gamma$ is spin wave damping 
factor and can also characterize the intrinsic magnon width, 
(1-e$^{-E/kT}$)$^{-1}$ is the Bose factor. 
In the small-\textbf{q} limit, the spin waves around (220) zone center can be approximately described by an anisotropic linear dispersion relation \cite{Garlea2008,Ramazanoglu2013}:

  \begin{equation}
       \hbar \omega(\textbf{q})=\sqrt{\Delta^{2}+\nu^{2}_{ab} (q^{2}_{x}+q^{2}_{y})}     
     \label{eq:dispersion}      
\end{equation}
    where $\nu$ is the spin-wave velocity and $\Delta$ is the energy gap. The constant-\textbf{Q} energy scans have been fitted 
using Eqs. (1) and (2) after convolution with the instrumental resolution using the RESLIB
program \cite {Zheludev}.
         
      The temperature dependence of the energy gap at (220) zone center is shown in Fig.\ \ref {fig:Escan} (b). Compared with the behavior of the 
energy gap in MnV$_{2}$O$_{4}$ \cite{Garlea2008} where only V has  orbital degrees of freedom, FeV$_{2}$O$_{4}$ shows three main differences: 1). 
In FeV$_{2}$O$_{4}$ the gap emerges below $T_{N1}$ whereas for MnV$_{2}$O$_{4}$ it only emerges below $T_{N2}$ ( = 53 K); 2). The gap in FeV$_{2}$O$_{4}$ is much higher; 3). No obvious increase in the 
energy gap ($\approx 1.5$ meV) as observed in MnV$_{2}$O$_{4}$ is found in FeV$_{2}$O$_{4}$ below $T_{N2}$ . 
     
      In the temperature range $T_{N2}<T<T_{N1}$ for both MnV$_{2}$O$_{4}$ and FeV$_{2}$O$_{4}$, the magnetic structures are similar without OO of
 V$^{3+}$.\cite{Nii2012}  The main difference in this temperature region between these two systems is that Fe$^{2+}$ is orbital-ordered whereas Mn$^{2+}$ is not. Thus, 
the appearance of an energy gap below $T_{N1}$ in FeV$_{2}$O$_{4}$ is due to the involvement of Fe$^{2+}$ OO  and not related
to V$^{3+}$ ions. It has been shown that the sole PM-CFI magnetic transition without any OO cannot induce an energy gap in MnV$_{2}$O$_{4}$\cite{Garlea2008}. Furthermore, the sole ferroic Fe$^{2+}$ 3\textit{z}$^{2}$-\textit{r}$^{2 }$ OO \cite{Nii2012} does not
 induce the energy gap below $T_{N1}$ since the OO is formed at a higher temperature $T_{S}$. Therefore, the spin-orbit coupling induced anisotropy at the A-site Fe$^{2+}$ is responsible for 
the appearance of the gap. High-resolution synchrotron x-ray measurements on FeV$_{2}$O$_{4}$ \cite {Nii2012} have shown that there is a strong spin-orbit coupling at the A-site 
Fe$^{2+}$ and the CFI ordering below $T_{N1}$ triggers the structural 
transition to orthorhombic with a lower symmetry via      
it,  similar to $A$Cr$_{2}$O$_{4}$ ($A=$ Fe, Cu) with 
only orbital-active ion at the A site \cite{Bordacs2009}. Compared with 
MnV$_{2}$O$_{4}$, the much higher  energy gap in 
FeV$_{2}$O$_{4}$  results from stronger spin-orbit coupling at Fe$^{2+}$ A 
site than that of the V$^{3+}$ B site in MnV$_{2}$O$_{4}$. The larger ordered moment of 4.0 $\mu$ $_{B}$ \cite{MacDougall2012} of Fe$^{2+}$ below $T_{N1}$
than that of the V ion (1.3 $\mu$ $_{B}$) in MnV$_{2}$O$_{4}$ \cite {Garlea2008}  also contributes to the higher gap. Below $T_{N2}$, although spin ordering of V$^{3+}$ are similar and V$^{3+}$ becomes orbital-ordered in both MnV$_{2}$O$_{4}$ and FeV$_{2}$O$_{4}$, the 
absence of measurable increase in energy gap below $T_{N2}$ in FeV$_{2}$O$_{4}$  implies nearly quenched orbital moments or a very weak SO coupling for the V$^{3+}$,  consistent with soft x-ray magnetic circular dichroism experiments \cite {Kang2012} and theoretical calculations \cite{Sarkar2011}. 
      
\begin{figure} \centering \includegraphics [width = 0.7\linewidth] {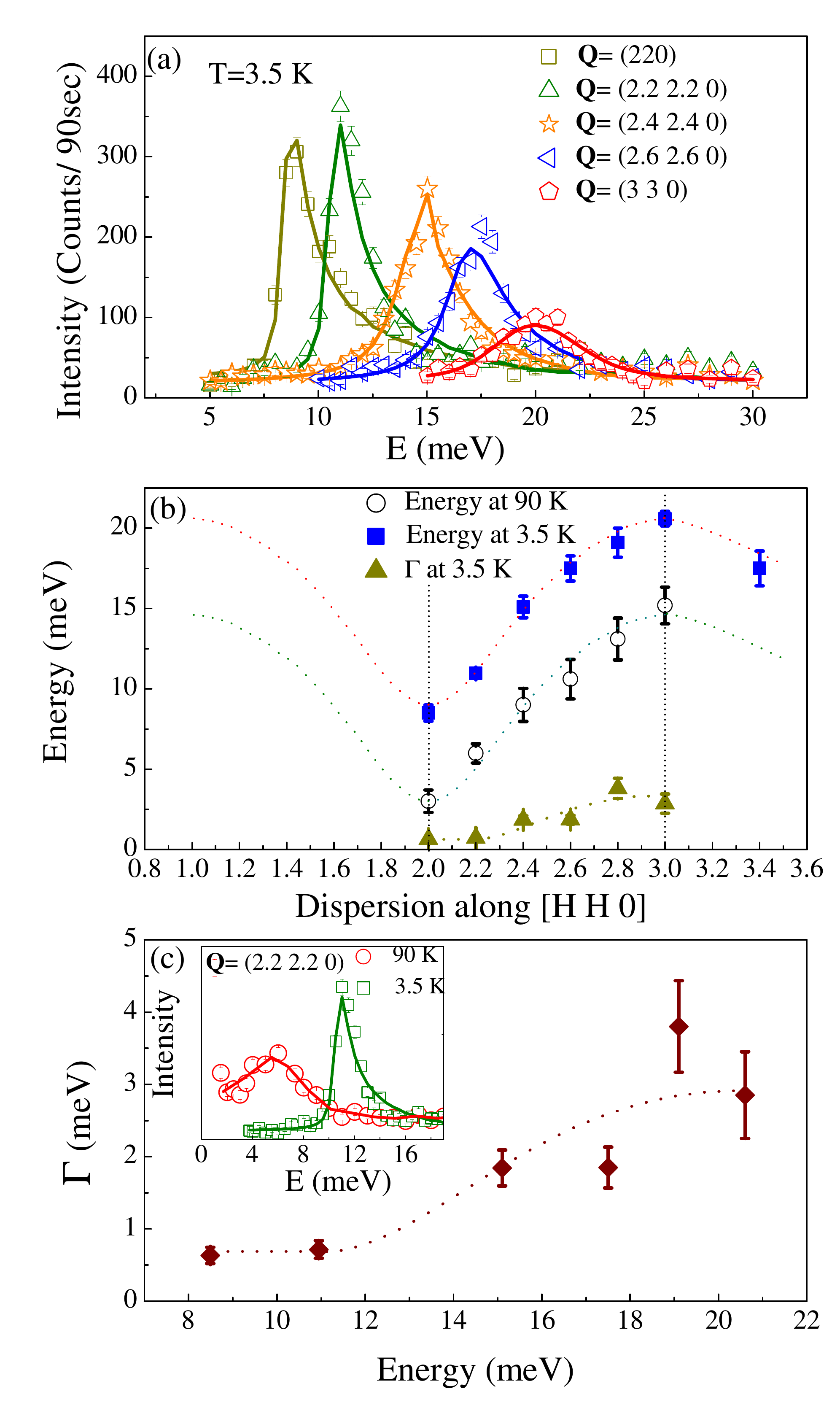}
\caption{(color online) (a) Constant-\textbf{Q} \textit{E }\textbf{}scans at different \textbf{Q}'s along [H H 0]  at 3.5 K in FeV$_{2}$O$_{4}$.
(b) 
Low-energy Fe spin waves with dispersion along [H H 0]  at 90 and 3.5 K, and the wave-vector
dependence of spin wave damping $\Gamma$ at 3.5 K . (c) Energy dependence of $\Gamma$ at 3.5 K . The dotted lines
are guides to the eye. The inset
shows a comparison of the raw data at \textbf{Q}=(2.2 2.2 0) at 90 and 3.5 K. }
\label{fig:SW} 
\end{figure} 

     To further get insight into the temperature evolution of the energy gap, we performed a least-square fit using power law $\Delta(T)\propto(T_{N1}-T)^{d}$ and obtained 
an exponent \textit{d} $\approx$ 0.75 below $T_{N1}$, similar to the value of $\approx$ 0.73 in MnV$_{2}$O$_{4}$ below $T_{N2}$.\cite{Garlea2008} This indicates the 
the energy gap induced by the anisotropy at Fe$^{2+}$ site in FeV$_{2}$O$_{4}$ has similar temperature evolution and critical behavior as the 
energy gap due to the anisotropy
 at V$^{3+}$ B site in MnV$_{2}$O$_{4}$. We also modeled the temperature dependence of the integrated intensity of Bragg peak (220) around  
$T_{N1}$ and $T_{N2}$ using  $I(T)\propto(T_{N1,N2}-T)^{2\beta}$,
 yielding critical exponents $\beta_{1}\approx 0.353$ at $T_{N1}$ and $\beta_{2}\approx 0.381$ at $T_{N2}$.  This indicates that FeV$_{2}$O$_{4}$
 is close to the 3D Heisenberg ($\beta=0.36$) or 3D Ising ($\beta=0.33$) models.
These two critical exponents are similar to the values near $T_{N1}$ and $T_{N2}$ reported in MnV$_{2}$O$_{4}$ \cite{Garlea2008}.
The comparison of $\beta_{1}$ with \textit{d} values below $T_{N1}$ in FeV$_{2}$O$_{4}$ indicates
that the temperature dependence of the energy gap varies like the square of the staggered magnetization \cite{Garlea2008} once the CFI ordering sets in.
     
   We emphasize that the spin wave damping factor $\Gamma$ 
(see Fig.\ \ref {fig:Escan} (b)) increases rapidly at $T_{N2}$. We also
used the Lorentz function convoluted with spectrometer resolution function 
to model constant-\textbf{Q} energy scans yielding FWHM values very close to $\Gamma$.
 Raising the temperature usually leads to a gradual spin wave broadening but the 
clear anomaly
 at $T_{N2}$ excludes the thermal effect only. We argue that the spin wave broadening in 
 $T_{N2}<T<T_{N1}$ originates from strong  
 fluctuations of V$^{3+}$ spins on the B site in the CFI phase prior to their true canting
  below $T_{N2}$. Such spin fluctuations of V$^{3+}$ provide interactions with the Fe spin waves and therefore lead to the anomalous spin wave broadening in $T_{N2}<T<T_{N1}$.
 
Representative constant-\textbf{Q} $E$ scans at various $Q$'s  along the [H H 0]  in the NCFI/ferroelectric phase are shown 
in Fig.\ \ref{fig:SW} (a). The solid lines are the best fit using the model described above and 
the obtained spin waves dispersions at the two different magnetic states are shown in Fig.\ \ref{fig:SW} (b).
Similar to the behavior at the (220) zone center,  $\Gamma$ at each fixed \textbf{Q}  at 90 K
is significantly larger than that at 3.5 K, as shown in the inset of Fig.\ \ref{fig:SW} (c). Compared with the spin-wave spectra of MnV$_{2}$O$_{4}$ \cite{Chung2008,Garlea2008}, the symmetric lowest-energy spin wave should be the acoustic 
mode due to the oscillations of Fe spins. 
Further measurements on other spin wave branches, especially four branches of V spin waves \cite{Chung2008} at higher energies are necessary to obtain accurate magnetic interactions 
such as \textit{J}$_{\rm Fe-V}$,
in-plane \textit{J}$_{\rm V-V}$, out-of-plane \textit{J}$_{\rm V-V}$, \textit{J}$_{\rm Fe-Fe}$ and the single-ion anisotropies. 
Note that the spin-wave shapes at 90 K and 3.5 K are very similar 
with only a shift of $\approx$ 5meV indicating that Fe$^{2+}$ spins are not influenced obviously
below $T_{N2}$.This suggests that there is no significant spin frustration at the Fe$^{2+}$ site in the NCFI/ferroelectric phase, consistent with powder 
neutron diffraction results that the direction of the Fe$^{2+}$ spins remain along the \textit{c} axis below/above $T_{N2}$. 
Therefore, the Fe spins without magnetic frustration at the A site are not mainly responsible for the appearance of the ferroelectricity below $T_{N2}$. 
Instead, the competition between the AFM \textit{J}$_{\rm V-V}$ and AFM \textit{J}$_{\rm Fe-V}$ induces strong spin frustration at the V site resulting in the canting of V$^{3+}$ spins,
 which plays a  major role in the appearance of the ferroelectricity based on (extended) spin-current models \cite{Katsura2005,Kaplan2011}.
   
   As shown in Fig.\ \ref{fig:SW} (b), the spin waves in FeV$_{2}$O$_{4}$ exhibit a significant broadening but without softening at the zone boundary. 
$\Gamma$ $\approx 0.7$  meV when H $\leq$ 2.2, but shows a significant 
increase/step to $\approx$ 3 meV above $H \approx 2.4$.
In Fig.\ \ref{fig:SW}(c), the average $\Gamma/E$ ratio
 $\approx$ 0.11 is much smaller than 
that of the metallic ferromagnetic La$_{2-2x}$Sr$_{1+2x}$Mn$_{2}$O$_{7}$ with 
$\Gamma/E \approx 0.33-0.46$ \cite{Perring2001}, consistent with the high insulating behavior of FeV$_{2}$O$_{4}$\cite{Zhang2012}. Furthermore, the $\Gamma/E$ $\approx 0.07$ at the zone center and $\approx 0.14$ at 
the zone boundary exhibit weak \textit{q} sensitivity of $\Gamma/E$ and $\Gamma$ is not linear with respect to $E$. All these features 
exclude magnon-electron 
scattering \cite{Perring2001} as the main mechanism for spin wave broadening. Theoretical calculations \cite {Furukawa1999} have shown that magnon-phonon coupling can increase the spin wave damping without softening the 
dispersion. Dai \textit{et. al.} \cite{Dai2000} reported a significant spin wave broadening at the zone boundary with a step in $\Gamma$  in a few manganese perovskites and demonstrated the role of magnon-phonon coupling as a mechanism of such anomalous
broadening. The existence of a strong coupling between spin, orbital, and lattice degrees of freedom \cite {Zhang2012} and the step in $\Gamma$ in FeV$_{2}$O$_{4}$ imply that  magnon-phonon coupling plays the main role in the spin wave broadening without softening at the zone boundary.

         In summary, neutron scattering studies on FeV$_{2}$O$_{4}$ crystal reveal the different roles of two 
orbital-active Fe$^{2+}$ at the A site and V$^{3+}$ at the B site in the magnetic excitations. 
The strong spin-orbit coupling at Fe$^{2+}$ A site induces a significant energy gap below $T_{N1}$ with little contribution from the V$^{3+}$.
The absence of a change in energy gap below $T_{N2}$ is evidence for a very weak SO coupling or significantly quenched
orbital moment of the V$^{3+}$. 
 Comparing the Fe spin waves below and above $T_{N2}$ precludes significant 
spin frustration at the Fe$^{2+}$ site, indicating Fe$^{2+}$ may not play an important role in inducing ferroelectricity. The important role of orbital-active Fe$^{2+}$ at the A site on the magnetic excitations is expected to be applicable to other spinels with orbital-active ions at that site, such as
$A$Cr$_{2}$O$_{4}$ ($A=$ Fe$^{2+}$ or Cu$^{2+}$). The anomalous spin wave broadening is observed in the collinear ferrimagnetic phase indicative of a possible 
V$^{3+}$ spin fluctuations prior to their true canting in the noncollinear ferrimagnetic phase. The spin wave broadening also exists at the zone boundary without 
obvious spin wave softening due to magnon-phonon coupling.

\section{Acknowledgments}  
      Research at Ames Laboratory is supported by the US Department of Energy, Office of Basic Energy Sciences, Division
of Materials Sciences and Engineering under Contract No. DE-AC02-07CH11358. Use of the high flux isotope reactor at the Oak Ridge National Laboratory,
 was supported by the US Department of Energy, Office of Science, Office of Basic
Energy Sciences, under Contract No. DE-AC02-06CH11357.

\textbf{SUPPLEMENTAL MATERIAL}\begin{LARGE}                                  \end{LARGE}

\subsection{Experimental details}
The FeV$_{2}$O$_{4}$ crystal was grown using the floating zone method. The DC susceptibility measurements were carried out 
on a Magnetic Property Measurement System (Quantum Design,SQUID). A big piece of 
crystal with the mass $\approx$ 1 g was cut for the elastic and inelastic 
neutron-scattering measurements that
 were conducted on the
 HB3 spectrometer (located at the High Flux Isotope Reactor, HFIR, at Oak Ridge National Laboratory, USA) with a fixed-final-energy ($E=14.7$ meV).
\subsection{Structural transitions in FeV$_{2}$O$_{4}$}

\begin{figure} \centering \includegraphics [width = 1\linewidth] {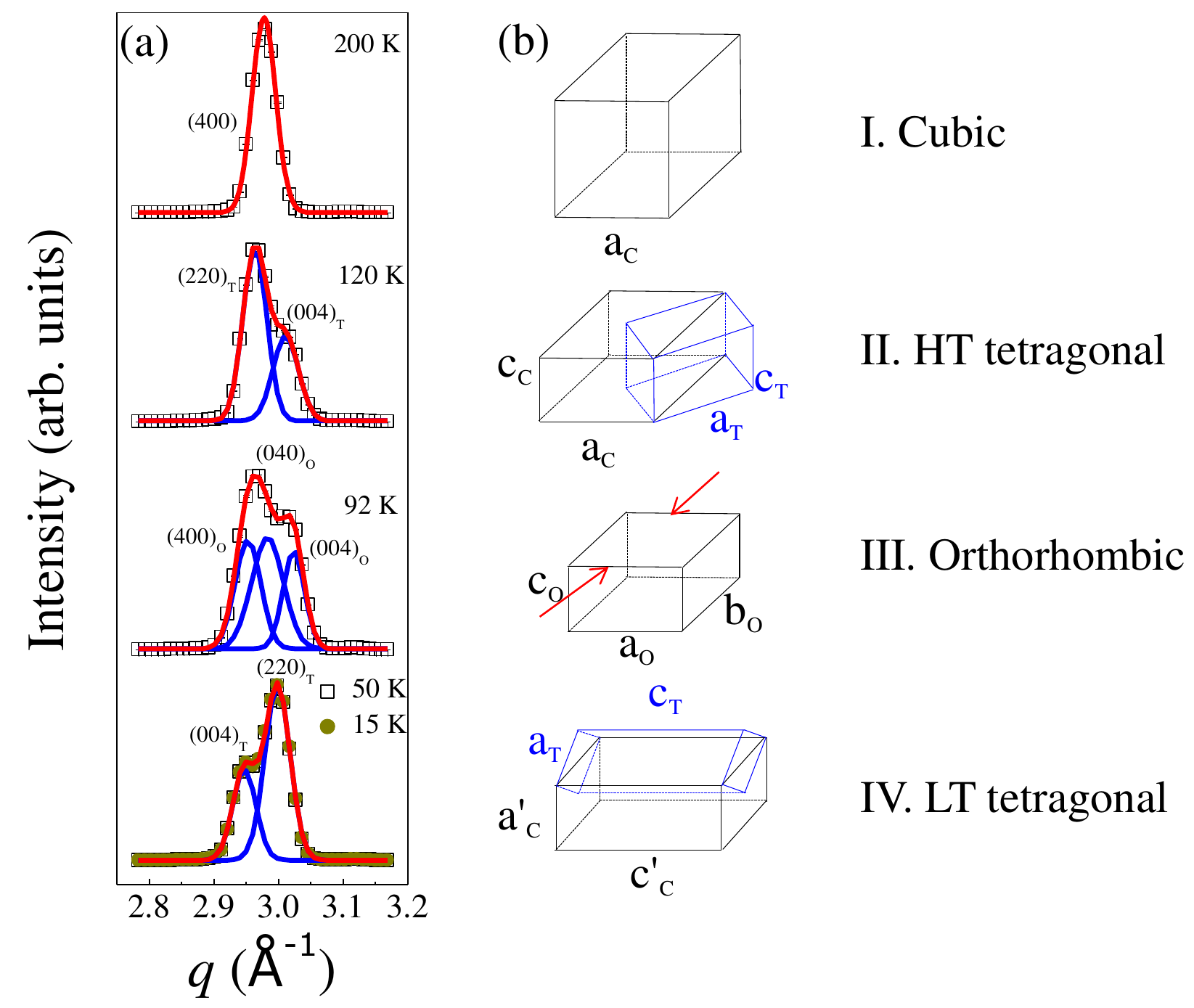}
\caption{(color online) 
(a).  Splitting of the (400) peak in the cubic setting of FeV$_{2}$O$_{4}$ at representative temperatures. Symbols are experimental data; red lines are the sum of two/three Gaussian fits; blue lines are the single Gaussian fit. 
(b) Schematic pictures of distortion of the crystals and the definition of directions of crystalline axes in different phases. Note that in both HT and LT tetragonal phases, the unit cell in the tetragonal
settting is one half of that distorted from the cubic unit cell above $T_{S}$. }
\label{fig:transitions} 
\end{figure}

        Figure\ \ref{fig:transitions} (a) shows the splitting of the \textit{q }scans of (400) structural Bragg peak in the cubic setting at several representative temperatures. 
 The schematic pictures of distortion of the crystals and the definition of directions of crystalline axes using different settings in different structures are summarized
 in Fig.\ \ref{fig:transitions} (b). At 120 K, one (400) peak splits to (220)$_{T}$ and (400)$_{T}$ in the tetragonal notation, suggesting the \textit{c} axis of the cubic unit cell is compressed due to Jahn-Teller distortion of FeO$_{4}$\cite{Katsufuji2008,Zhang2012,MacDougall2012} driven by a ferroic Fe$^{2+}$ 3\textit{z}$^{2}$-\textit{r}$^{2 }$ OO\cite{Nii2012} and the cubic structure transforms into 
 high-temperature (HT) tetragonal structure with $ c_{c}<a_{c}$  ($a_{c}$  is the lattice constant in the cubic phase).
As temperature further decreases to 92 K, one of the \textit{a} axis in the HT tetragonal phase
 is compressed\cite{Katsufuji2008}, resulting in a structural transition to 
orthorhombic phase, which can be seen 
 from the splitting of two peaks to three peaks (400)$_{\rm O}$, (040)$_{\rm O}$ 
and (004 )$_{\rm O}$.
At 50 K, these three peaks evolve back to two peaks, indicating another structural transition to low-temperature (LT) tetragonal phase. As opposed to the HT tetragonal phase,
the peak position \textit{q} of (004)$_{T}$ is smaller than that of (220)$_{T}$ peak in the LT tetragonal phase. 
Moreover, the peak intensity of low-$q$ peak is weaker than that of 
the high-q peak in the LT tetragonal phase. 
Thus, the LT tetragonal 
phase with $c>a$ is different from the HT tetragonal phase with 
$c<a$ and the tetragonal \textit{c} axes in these two 
tetragonal phases are perpendicular to each 
other in one unit cell. This indicates that the compressed axis 
(\textit{b$_{\rm O}$} or \textit{a$_{\rm O}$}) in the orthorhombic phase, 
becomes equal to the \textit{c$_{\rm O}$}, and therefore, the orthorhombic phase 
evolves to LT tetragonal phase and the remained axis becomes new \textit{c} axis in the tetragonal setting. Based on the above discussion, we notice that during the structural transformation from HT tetragonal phase to
LT tetragonal phase in FeV$_{2}$O$_{4}$, the orthorhombic phase cannot be avoided. It should be noted that as shown in Fig. 1 (a), no  change is 
observed in the \textit{q }scans of (400) structural Bragg peak between 50 K and 15 K, excluding any structural transition at around 35 K as reported by Katsufuji \textit{et al.}\cite{Katsufuji2008}.
 
      It is worthwhile noting that the HT tetragonal-orthorhombic structural transition, accompanied by the paramagnetic-collinear ferrimagnetic transition \cite{MacDougall2012} occurs at 70 K here,
consistent with the value in \textit{Ref.} \cite{Katsufuji2008}, is a little higher than the value of 56 K in the polycrystalline sample \cite{Zhang2012,MacDougall2012} and also 
the value of 60 K \cite{MacDougall2012} or 65 K \cite{Nii2012} in the single crystal form, 
reflecting strong suppression of the 
nonstoichiometry, i.e., $x\approx0$ in the formulation (Fe$^{2+}$)(Fe$^{3+}_{x}$V$^{3+}_{2-x}$)O$_{4}$ in our crystal since a lower value of $x$ causes an increase of $T_{N2}$ \cite{Liu2012}.
Thus, the investigation on the stoichiometric FeV$_{2}$O$_{4}$ crystal can minimize the effect of nonstoichiometry and reveal intrinsic magnetic excitations.

\end{document}